\documentclass{article}

\usepackage{alltt}
\usepackage{amsmath}
\usepackage[title]{appendix}
\usepackage{authblk}
\usepackage{bm}
\usepackage{color}
\usepackage{graphicx}
\usepackage{hyperref}
\usepackage{mathtools}
\usepackage{natbib}
\usepackage{placeins}
\usepackage{setspace}
\usepackage{subcaption}

\let\proglang=\textit
\let\code=\texttt

\begin{document}

\doublespacing

\title{ How big does a population need to be before demographers can ignore individual-level randomness in demographic events? }

\author[1]{John Bryant}
\author[2]{Tahu Kukutai}
\author[3]{Junni L. Zhang}

\affil[1]{Bayesian Demography Limited, Christchurch, New Zealand}
\affil[2]{Te Ngira Institute for Population Research, University of Waikato, Hamilton, New Zealand}
\affil[3]{National School of Development, Peking University, Beijing, China}

\date{}

\maketitle

\begin{abstract}
BACKGROUND
When studying a national-level population, demographers can safely ignore the effect of individual-level randomness on age-sex structure. When studying a single community, or group of communities, however, the potential importance of individual-level randomness is less clear.

OBJECTIVE
We seek to measure the effect of individual-level randomness in births and deaths on standard summary indicators of age-sex structure, for populations of different sizes. We focus on demographic conditions typical of historical populations.

METHODS
We conduct a microsimulation experiment where we simulate events and age-sex structure under a range of settings for demographic rates and population size.

RESULTS
Individual-level randomness strongly affects age-sex structure for populations of about 100, but has a much smaller effect on populations of 1,000, and a negligible effect on populations of 10,000.

CONCLUSIONS
Analyses of age-sex structure in historical populations with sizes on the order 100 must account for individual-level randomness in demographic events. Analyses of populations with sizes on the order of 1,000 may need to make some allowance for individual-level variation, but other issues, such as measurement error, probably deserve more attention. Analyses of populations of 10,000 can safely ignore individual-level variation.

CONTRIBUTION
We provide rules of thumb for when it is safe to ignore individual-level randomness in demographic events when analysing age-sex structure in small historical populations.

\end{abstract}

\section{Introduction}

Demography has traditionally emphasised national populations, which are large enough that randomness in individual demographic events such as births and deaths averages out and can be ignored. In such populations, variation in population size and structure is almost entirely driven by variation in the underlying birth, death, and migration rates \citep[][p. 157]{lee1998probabilistic,preston2001demography}.

The study of demographic change can, however, often benefit from narrowing the focus down to individual communities or localities, and reconstructing events and processes at the micro level \citep{caldwell1987anthropology,caldwell1990cultural}. Local-level demographic studies have generated fundamental knowledge about historical patterns of survivorship and fertility \citep{henry1968historical}. Historical demography in France began with multiple studies of villages, using parish records to reconstitute families and their demographic behaviours \citep{rosenthal1997thirteen}. These studies, and later ones in England \citep{wrigley1989population} and continental Europe \citep{coale1986decline}, profoundly changed understanding about population change and its social and economic drivers. Local-level demographic studies can contribute to the understanding of social change elsewhere in the world, such as documenting the demographic effects of colonisation on Indigenous peoples \citep{cook1971essays,covey2011dynamics,petersen1975demographers}, where large-scale systematic demographic data are typically lacking \citep{dobyns1966estimating,roberts1989disease,thornton1980recent}.

If a population is small, ignoring individual-level randomness may be unwise. In a population of 100 people, for instance, the number of deaths can easily halve or double from one year to the next with no change in the underlying risk of dying. (For a demonstration, see Appendix~\ref{sec:halve_or_double}.) If deaths, births, or migrations vary randomly from year to year, we would expect population size and structure to vary randomly as well. A run of higher-than-average birth counts, for instance, should increase population size and produce a younger age structure. 

The relationship between individual-level randomness in demographic events and population-level randomness in size and structure is, however, complicated, because the relations between demographic events and population size and structure are subject to lags and feedback loops. The number of births, deaths, and migrations that occur this year, for instance, reflects the size and structure of the population this year, but the size and structure of the population this year is itself a product of births, deaths, and migrations during previous years.

The complexity of the relationship between births, deaths, migration, and population makes it difficult to know how much randomness to expect in small populations. Having reliable intuitions about random variation in small populations is, however, crucial to making inferences about the demography of these populations. Consider, for instance, a demographer who has data on the population share of old people in two communities of around 1,000 people. If 5\% of people in one community are aged 60+ and 10\% in the other are aged 60+, how confident should the demographer be that the two communities have different demographic regimes?

One way to build up intuitions about the amount of random variation to expect in small populations would be to assemble large quantities of accurate data on birth rates, death rates, and population size and structure from a wide variety of settings, and see how much variability there is across communities. However, for many populations,, particularly historical populations, such data are simply not available, so the direct observational approach is not feasible.

Another way to build intuitions is to use mathematical theory. \citet[][ch. 15]{caswell2001matrix} presents a sophisticated mathematical framework for studying the effects of individual-level variability on population aggregates.

We, however, take a more direct approach that requires only elementary mathematics, and hence is hopefully more accessible to applied demographers. Our approach is computer simulation.  We randomly generate artificial life histories which we aggregate up into populations classified by age and sex.  The strategy of using computer power as an alternative to mathematical theory is common in contemporary applied statistics \citep{gelman2021most}, though there is also a long tradition in demography of using micro-simulation to study demographic processes \citep{wachter1978statistical,van1998microsimulation,zagheni2015microsimulation}.
 
Our simulations use three different initial population sizes: 100, 1,000, and 10,000. Having different population sizes allows us to examine the relationship between population size and random variability. We also simulate a variety of different mortality and fertility regimes. This gives us a yardstick for assessing the importance of randomness: we can compare the variation attributable to randomness in births and deaths with the variation attributable to differences in underlying demographic rates. For simplicity, we assume zero migration, and we assume that the underlying age-specific birth rates and death rates are constant throughout the simulation period.

The study is designed to mimic the demographic conditions that are studied by historical demographers, with high mortality and high fertility. We also assume that, due to limitations in available data, demographers must work with broad indicators of population structure, such as dependency ratios.

Our simulations suggest that, in populations of around 100 people, variation due to the randomness of demographic events is large relative to the variation due to differences in underlying rates. When population size reaches 1,000, however, variation due to the randomness of demographic events is already relatively modest. By the time population size reaches 10,000, variation due to the randomness of demographic events is negligible.

All data and code for the simulations are available at [names of github repositories].

\section{Design of the simulations}

\subsection{Inputs}

The simulations require values for  population size, mortality, the sex ratio at birth, and fertility. The inputs are summarized in Table~\ref{tab:inputs}.

\begin{table}[h]
  \centering
  \renewcommand{\arraystretch}{1.2}  
  \caption{Inputs to the simulation}
  \begin{tabular}{ l p{8cm} }
    \hline
    Input & Description \\
    \hline
    $N$ & Initial population size, set by us \\
    $m_{as}$, $L_{as}$ & Mortality rates and expected person-years lived, from West model life tables \\
    $p_s$ & Proportion of births of sex $s$, based on a sex ratio at birth of 105 \\
    $g_a$ & Age-specific fertility rates, scaled to sum to 1, from the `Booth standard' \\
    $F$ & The total fertility rate, supplied by us. \\
   \hline                         
  \end{tabular}
  \label{tab:inputs}
\end{table}

Population size $N$ is the total number of people in the population at the start of the simulation. Values for $N$ are chosen by us to be 100, 1,000, and 10,000. 

The $m_{as}$ in the second row of Table~\ref{tab:inputs} is the mortality rate for people of sex $s$ in age group $a$. All age groups have widths of 5 years, apart from the final age group, which includes everyone aged 95 years and over. The quantity $L_{as}$ is the number of years that a newborn baby of sex $s$ could expect to live in age group $a$, under prevailing mortality rates.

We obtain values for $m_{as}$ and $L_{as}$ from West model life tables \citep{coale1983models}, as reported in the \proglang{R} package \code{demogR} \citep{jones2007demogr}. The West model life tables are a set of mortality rates, and derived quantities, that demographers use to represent typical mortality patterns across a wide range of settings. West model life tables are indexed by level, with lower lowers indicating higher mortality rates, and hence lower life expectancies. Table~\ref{tab:life_exp} shows the West model life table levels that we use in the simulations, along with the associated life expectancies. These life tables represent mortality conditions in pre-industrial societies.

\begin{table}[ht]
\centering
\caption{Life expectancy for each level of the West model life tables used in the simulations} 
\label{tab:life_exp}
\begin{tabular}{rrrr}
  \hline
 & Level 1 & Level 5 & Level 9 \\ 
  \hline
Female life expectancy & 20.0 & 30.0 & 40.0 \\ 
  Male life expectancy & 18.0 & 27.7 & 37.3 \\ 
   \hline
\end{tabular}
\end{table}

We set $p_{\text{F}}$, the proportion of births that are female, to 0.488. This is equivalent to assuming that there are 105 male births per 100 female births, which is a standard assumption in demography. 

The `Booth standard' age pattern for fertility describes a typical distribution of fertility across women's reproductive years. It is the starting point for some widely-used methods for fertility estimation \citep{booth1984transforming, moultrie2013tools}. Values for the Booth standard are shown in Table~\ref{tab:booth}. 

\begin{table}[ht]
\centering
\caption{Proportional distribution of births by age of mother, under the Booth standard} 
\label{tab:booth}
\begin{tabular}{rrrrrrrr}
  \hline
10--14 & 15--19 & 20--24 & 25--29 & 30--34 & 35--39 & 40--44 & 45--49 \\ 
  \hline
0.003 & 0.133 & 0.241 & 0.231 & 0.188 & 0.134 & 0.062 & 0.008 \\ 
   \hline
\end{tabular}
\end{table}

The total fertility rate (TFR) is the average number of births a woman could expect to have over her life if she survived to the end of the reproductive ages, under prevailing fertility rates. Values for the TFR are chosen by us, and are 4, 5, and 6.

\begin{figure}
  \centering
     \includegraphics{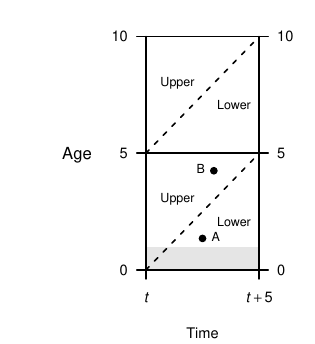}
        \caption{\small Disaggregating events by Lexis triangle. Event A belongs to the lower Lexis triangle for age group 0--4, while event B belongs to the upper Lexis triangle for age group 0--4. The grey area is in children aged less than 1 year.}
      \label{fig:lexis}
\end{figure}

To connect deaths and births to population, classifying deaths and births by age, sex, and period is not sufficient: we also need to classify them by `Lexis triangle'.  Figure~\ref{fig:lexis} depicts Lexis triangles for deaths for age groups 0--4 and 5--9. The triangles are defined by the solid lines marking out age groups and periods, and the dashed lines on the diagonal.  A death for age group $a$ belongs to the upper Lexis triangle if the person dying was in age group $a$ at the start of the period; otherwise the death belongs to the lower Lexis triangle.

With fertility, we assume that, for any given age, the rate in the lower Lexis triangle is identical to the rate in the upper Lexis triangle.  For mortality we make the equivalent assumption for all age groups except the first. For ages 0--4, assuming equality is too crude an approximation. The mortality rate experienced by infants is much higher than the mortality rate experienced by children aged 1--4, and values in the lower and upper Lexis triangles need to be weighted accordingly. Appendix~\ref{sec:lexis_child} describes our method for doing so.

\subsection{Initial populations}

Choosing an appropriate initial age-sex structure for the simulations is not trivial. Unlike populations in which births and deaths are deterministic, populations in which births and deaths are random do not have straightforward stable population structures that might serve as a neutral starting point \citep[][ch. 15]{caswell2001matrix}. We instead use the stationary populations implied by the associated life tables in a deterministic model. We start with the age structure that would arise in a deterministic setting where births equalled deaths, in which the share of the population in age group $a$ and sex $s$ is proportional to $L_{as}$ \citep[][ch. 5]{keyfitz2005applied}. We then scale this population so that it equals $N$ and then round each cell to the nearest integer.

\subsection{Simulating death and survival}

As a cohort ages, it moves diagonally across the Lexis diagram, alternating between Lower and Upper triangles. We simulate death and survival one Lexis triangle at a time. Within each Lexis triangle, we simulate individuals. To simulate an individual, we go through the following steps:
\begin{enumerate}
  \item Draw a value for the amount of time left before the individual exits the current Lexis triangle (which for a lower Lexis triangle means reaching time $t+5$ and for an upper Lexis triangle means attaining age $a+5$.) We assume that, at the start of each Lexis triangle, individuals are spaced evenly across the time axis or the age axis, which allows us to derive the amount of time left by drawing from a uniform distribution with minimum 0 and maximum 5.
  \item Draw a value for the amount of time before the individual dies, given the mortality rates in the Lexis triangle. This is done by drawing from an exponential distribution with the associated rate.
  \item If the second value is shorter than first, increment the number of deaths for that Lexis triangle by one, and remove the individual from the simulated population; otherwise include the individual in the population entering the next Lexis triangle for that cohort.
\end{enumerate}

\subsection{Simulating births}

Births are also simulated one triangle at a time. If the individual being simulated is a female in the reproductive ages, then we multiply the amount of time that the individual spent in the Lexis triangle by the fertility rate for that triangle, to obtain an expected number of births, and then draw from a Poisson distribution with that expected number. These births are added to the population entering the lower Lexis triangle for age group 0--4.

\subsection{Summary measures of population structure}

We examine the effect of individual-level randomness on four summary measures of population structure. We use these particular indicators on the grounds that they are of substantive interest in a wide variety of demographic studies, and can often be calculated from the type of data that is available for historical populations, such as counts of children and adults. The indicators are as follows:

\begin{itemize}
  \item Percent of population aged 0--14
  \item Percent of population aged 60+
  \item Child-woman ratio: The ratio of children aged 0--14 to women aged 15+
  \item Dependency ratio: The ratio of people aged 0--14 and 60+ to people aged 15--59.
\end{itemize}

\subsection{Implementation}

The data preparation and the analysis of results is done in R, while the simulation code is written in C++, for speed. We do 10,000 simulation runs for each combination of inputs. The total time required for all simulation runs combined is around 15 minutes. As noted above, all the code is available at [addresses of github repositories].

\subsection{Alternative approach to simulation}

We also constructed simulations using the \texttt{estimateAccount} function in R package \code{demest}\footnote{github.com/statisticsnz/demest}. The results were very similar to those from the microsimulation, but the calculations took 20 hours rather than 15 minutes.

\section{Results}

\subsection{Four simulated populations}
  \label{subsec:four_popn}

\begin{figure}
  \centering
  \begin{subfigure}[b]{0.48\textwidth}
        \includegraphics{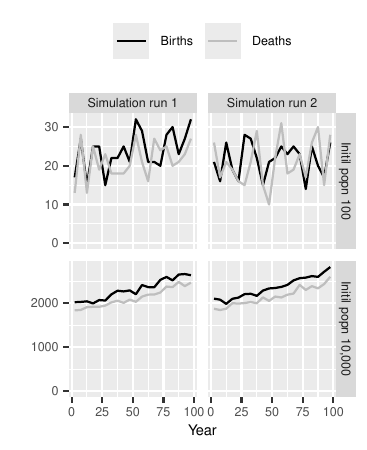}
        \caption{Birth and death counts for five-year periods.}
        \label{fig:example_birth_death}
  \end{subfigure}
  \begin{subfigure}[b]{0.48\textwidth}
        \includegraphics{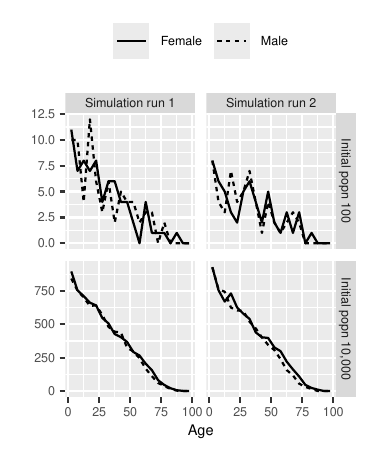}
        \caption{Population counts, by age and sex, at year 100}
        \label{fig:example_popn}
  \end{subfigure}
  \caption{Results from four randomly-chosen simulations generated using a West level 5 life table and a TFR of 5.}
      \label{fig:example}
\end{figure}

To give a sense of what the simulations contain, we first present output from four randomly-chosen simulation runs, generated using a West level 5 life table and a total fertility rate of 5. Figure~\ref{fig:example_birth_death} shows birth and death rates. The populations in the top row of Figure~\ref{fig:example_birth_death} were generated using initial population size $N =$~100, and the populations in the bottom row were generated using initial population size $N =$~10,000. The two small populations are much more variable over time than the two large populations, and look less similar to each other. Figure~\ref{fig:example_popn} shows age-sex structure at year 100. The small populations again differ much more than the large populations.

\subsection{Distribution of summary indicators for total fertility rate 5 and life table level 5}

The results from the four simulation runs shown in Figure~\ref{fig:example} are consistent with the idea that smaller population size is associated with more variability. With only four simulated populations, however, it is difficult to assess the strength of the relationship. To do this, we need more simulation runs.

\begin{figure}[h]
  \centering
  \includegraphics{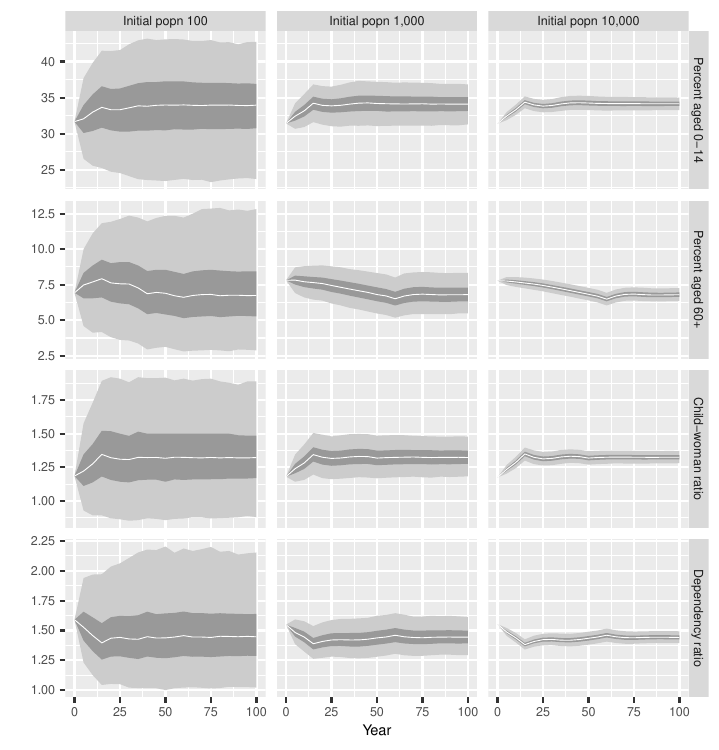}
  \caption{Distributions of summary indicators, over time, for populations with TFR of 5 and West life table level 5. The light grey bands represent the interval between the 2.5\% and 97.5\% quantiles; the dark grey bands represent the interval between the 25\% and 75\% quantiles; and the white lines represent medians.}
      \label{fig:time_5_5}
\end{figure}

Figure~\ref{fig:time_5_5} shows distributions of summary measures over 10,000 runs. Values for the summary measures evolve over time, so the figure shows values for each year of the simulations. In all simulation runs in Figure~\ref{fig:time_5_5}, the total fertility rate is 5 and the West life table level is 5. Each panel shows a different combination of summary indicator and initial population size. The top-left panel, for instance, shows the percent of the population aged 0--14 in a population with starting size 100.

 The white lines in Figure~\ref{fig:time_5_5} show median values across the indicator across all 10,000 simulations. The lower limit of the dark grey bands represents the 25\% quantile: 25\% of simulations, at each time point, have values for the indicator that are lower than this limit. The upper limit of the dark grey bands represent the 75\% quantile: 75\% of simulation runs have values that are lower than this limit. The light grey bands represent the interval between the 2.5\% and 97.5\% quantiles. The wider the dark grey and light grey bands, the more variability there is. 

For a given combination of settings, the only source of variability across populations is random variation in demographic events. The width of the grey bands therefore shows the impact on the population indicators of random variation in demographic events. Comparing widths across columns shows how sensitivity to random variation declines with population size. 

The key finding from Figure~\ref{fig:time_5_5} is that, with this particular set of parameters, variability falls sharply between populations with initial size $100$ and populations with initial size $1,000$, and falls less sharply between populations with initial size $1,000$ and initial size $10,000$. Populations of around $100$ can have vastly different values for the four summary indicators, even when their underlying fertility and mortality rates are the same. Populations of around $10,000$ show some differences in the summary indicators, but these differences are relatively small. Populations of around $1,000$ behave more like populations of $10,000$ than like populations of $10,000$.

\clearpage

\subsection{Distributions of summary indicators for all combinations of total fertility rate and life table level}
  \label{subsec:distributions}

\begin{figure}[h]
  \centering
     \includegraphics{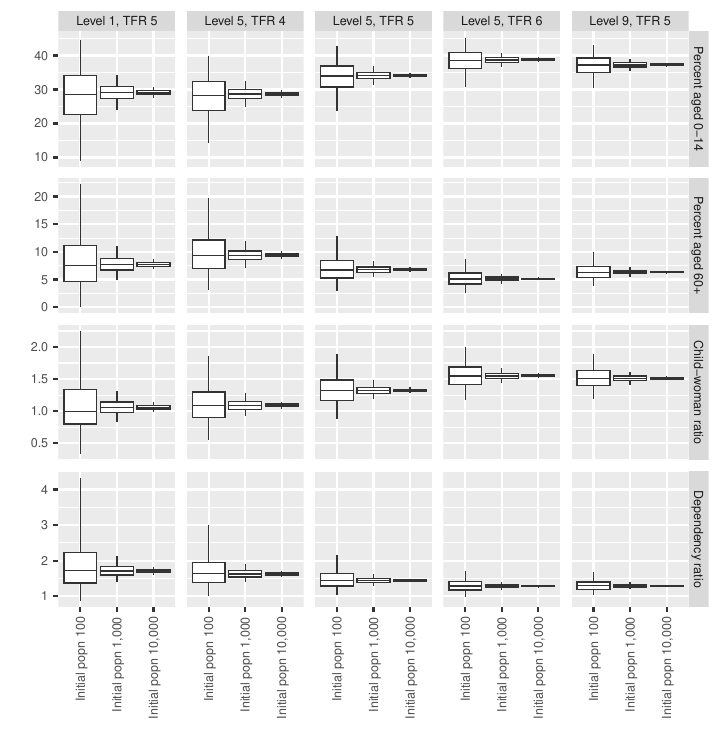}
        \caption{Distributions of summary indicators at year 100 of the simulations, for every combination of population size, fertility rates, and mortality rates. The tips of the vertical lines represent 2.5\% and 97.5\% quantiles; the tops and bottoms of the boxes represent 25\% and 75\% quantiles; and the horizontal lines represent medians.}
      \label{fig:summary}
\end{figure}

Figures~\ref{fig:time_1_5}--\ref{fig:time_9_5} in the Appendix show results equivalent to those of Figure~\ref{fig:time_5_5} for other combinations of TFR and West life table levels. Comparing across the five graphs is, however, difficult. To simplify comparisons, we restrict our attention to a single post-transition year. Figure~\ref{fig:summary} focuses on year 100, and gives results for five combinations of TFR and life table level. The tops and bottoms of the boxes in Figure~\ref{fig:summary} correspond to the 25\% and 75\% quantiles in Figure~\ref{fig:time_5_5}; the tops and bottoms of the vertical lines correspond to the tops and bottoms of the 2.5\% and 97.5\% quantiles in Figure~\ref{fig:time_5_5}; and the horizontal lines correspond to the medians.

In Figure~\ref{fig:summary}, like in Figure~\ref{fig:time_5_5}, there is a large decline in variability when moving from initial population size  $100$ to initial size population size $1,000$. There is also a decline in variability when moving from initial population size $1,000$ to initial population size $10,000$, but it is more modest.

With initial population size $100$, random variability in the indicators is sufficiently large, relative to differences across the five demographic scenarios, that the distributions largely overlap. The distribution of the percent aged 0--14 under a TFR of 5 and a life table level of 1, for instance,  overlaps with the distribution of the percent aged 0--14 under a TFR of 5 and a life table level of 5.

With initial population size $10,000$, random variability is much less, and the distributions of the indicators are accordingly much more distinct. There is little overlap, for instance, between the distributions generated by a TFR of 5 and a life table level of 1 and the distributions generated by a TFR of 5 and a life table level of 9.

The amount of overlap with initial population size $1,000$ is again intermediate between the other two sizes. However, the results are again closer to those for size $10,000$ than to size $100$.

\clearpage

\section{Discussion}
  \label{sec:discussion}

Our simulation results imply that, under the types of high-fertility, high-mortality regimes typical of historical populations, demographic analyses of very small populations, on the order of 100 people, need to be highly attentive to random variation.  It is risky to use differences in age structure to make inferences about differences in underlying demographic rates, because the differences in age structure could easily be due to random variation.

Demographic analyses of very small populations need to model this random variation. One way of doing so when data from multiple small populations is available is to use Bayesian hierarchical models that pool information across the populations, and that allow information about plausible ranges to be incorporated in a transparent and systematic way. \citet{schmertmann2019bayesian} is an example of this approach. A second, complementary, approach is to build population simulations that explicitly model the sources of variation. The simulation methods presented in this paper are one way of doing so.

Our simulation results also imply that once population sizes reach 1,000 or so, variation attributable to randomness in births and deaths has already reached modest levels, not much greater than would be expected for much larger populations. Demographers dealing with populations of this size do, of course, need to be careful not to over-interpret small differences in age-sex structure. But it appears that variability that is specifically due to randomness in births and deaths is less of an obstacle to demographic analyses of these populations than might be thought. Demographers working with broad measures of age-sex structure in historical populations would probably be better to devote scarce analytical resources to assessing the impact of issues such as measurement error or variation in underlying demographic rates than to individual-level random variation in demographic events.

Demographers working with broad measures of age-sex structure in populations of 10,000 or high can, like demographers working with national-level populations, safely ignore individual-level variation in demographic events.

\bibliographystyle{apalike}
\bibliography{main}

\begin{appendices}

\section{Year-to-year changes in observed death rates in populations}
  \label{sec:halve_or_double}

Here we use some simple simulations to support the claim from the Introduction that the number of deaths can easily half or double from one year in a small population.  We use a population size of 100, we assume that deaths follow a Poisson distribution, and we ignore the effect of the annual number of deaths on the annual population at risk, which is small in percentage terms. We simulate deaths in pairs of years, and calculate the proportion of pairs in which the ratio of deaths in one year to deaths in the other year is less than 0.5 or greater than 2.

\begin{verbatim}
percent_halve_or_double <- function(popn_size, death_rate) {
    ## Calculate expected number of births
    lambda <- popn_size * death_rate
    ## Generate 'n' births in first and second years
    deaths_first_year <- rpois(n = 1000000, lambda = lambda)
    deaths_second_year <- rpois(n = 1000000, lambda = lambda)
    ## Obtain the ratio of births in year 1 to births in year 2,
    ## excluding cases where births in both years are 0
    ratio <- deaths_second_year / deaths_first_year
    combined_deaths <- deaths_first_year + deaths_second_year
    ratio[combined_deaths == 0] <- 1
    ## Calculate the proportion of cases where births
    ## in year 2 where less than half, or more than twice,
    ## births in year 1
    ans <- mean((ratio < 0.5) | (ratio > 2))
    ## Convert to percent
    round(ans * 100)
}

set.seed(0)
percent_halve_or_double(popn_size = 100, death_rate = 0.02)
## [1] 44
percent_halve_or_double(popn_size = 100, death_rate = 0.03)
## [1] 37
percent_halve_or_double(popn_size = 100, death_rate = 0.04)
## [1] 31
\end{verbatim}

\section{Calculating mortality rates for Lexis triangles for age 0--4}
  \label{sec:lexis_child}
  
  We calculate life table mortality rates by Lexis triangle, for age group 0--4 and sex $s$, by first calculating person-years lived and deaths for each triangle. The total number of person-years lived within age group 0 over a 5-year period is $5 L_{0s}$.  The total number of deaths is $5 L_{\text{0} s} m_{\text{0} s}$. Similarly, the total number of person-years lived and deaths for age group 1--4 is $5 L_{\text{1-4},s}$ and $5 L_{\text{1-4},s} m_{\text{1-4},s}$. We allocate person-years lived according to the degree of overlap between each Lexis triangle and age group 0, depicted as a grey rectangle in Figure~\ref{fig:lexis}. The lower Lexis triangle thus receives 4.5/5 of the person-years lived within age group 0, and the upper Lexis triangle receives 0.5/5 person years. Deaths are allocated the same way. This yields the formulas
\begin{align}
  m_{\text{0--4},s,\text{Low}} & = \frac{\text{Deaths in lower triangle}}{\text{Person-years lived in lower triangle}} \\
  & = \frac{ (4.5/ 5) \times 5 L_{\text{0},s} m_{\text{0},s} + (8/20) \times 5 L_{\text{1--4},s} m_{\text{1--4},s} } 
  {  (4.5/ 5) \times 5 L_{\text{0},s} + (8/20) \times 5 L_{\text{1--4},s} }  \\
  & = \frac{ 4.5 L_{\text{0},s} m_{\text{0},s} + 2 L_{\text{1--4},s} m_{\text{1--4},s} } 
  {  4.5 L_{\text{0},s} + 2 L_{\text{1--4},s} } \\
  m_{\text{0--4},s,\text{Up}} & = \frac{\text{Deaths in upper triangle}}{\text{Person-years lived in upper triangle}} \\
  & = \frac{ (0.5/ 5) \times 5 L_{\text{0},s} m_{\text{0},s} + (12/20) \times 5 L_{\text{1--4},s} m_{\text{1--4},s} } 
  {  (0.5/ 5) \times 5 L_{\text{0},s} + (12/20) \times 5 L_{\text{1--4},s} }  \\
  & = \frac{ 0.5 L_{\text{0},s} m_{\text{0},s} + 3 L_{\text{1--4},s} m_{\text{1--4},s} } 
  {  0.5 L_{\text{0},s} + 3 L_{\text{1--4},s} }.
\end{align}

\section{Plots of simulations over time}
  \label{sec:time}

\begin{figure}[h]
  \centering
  \includegraphics{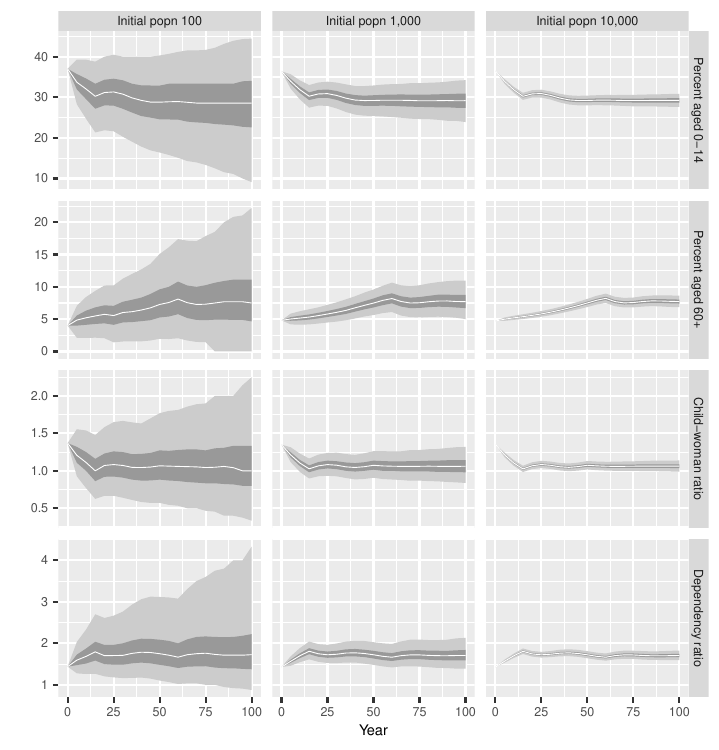}
  \caption{Distributions of summary indicators, over time, for populations with TFR 5 and West life table level 1.}
      \label{fig:time_1_5}
\end{figure}  

\begin{figure}[h]
  \centering
  \includegraphics{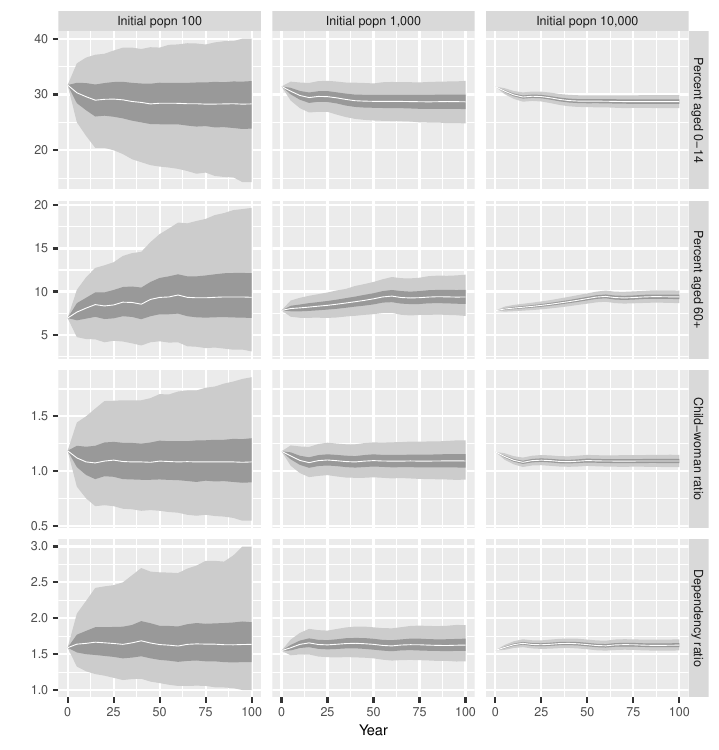}
  \caption{Distributions of summary indicators, over time, for populations with TFR of 4 and West life table level 5.}
      \label{fig:time_5_4}
\end{figure}  

\begin{figure}[h]
  \centering
  \includegraphics{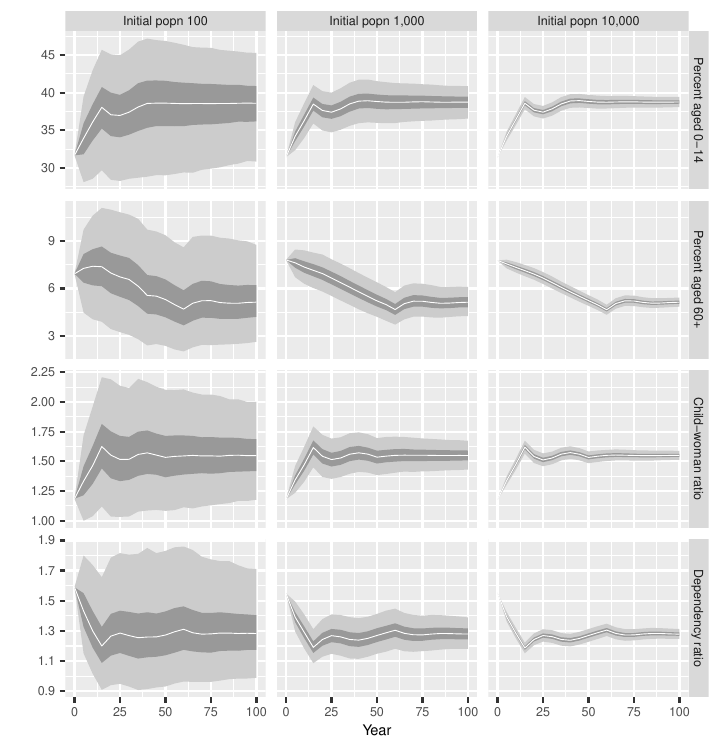}
  \caption{Distributions of summary indicators, over time, for populations with TFR of 6 and West life table level 5.}
      \label{fig:time_5_6}
\end{figure}

\begin{figure}[h]
  \centering
  \includegraphics{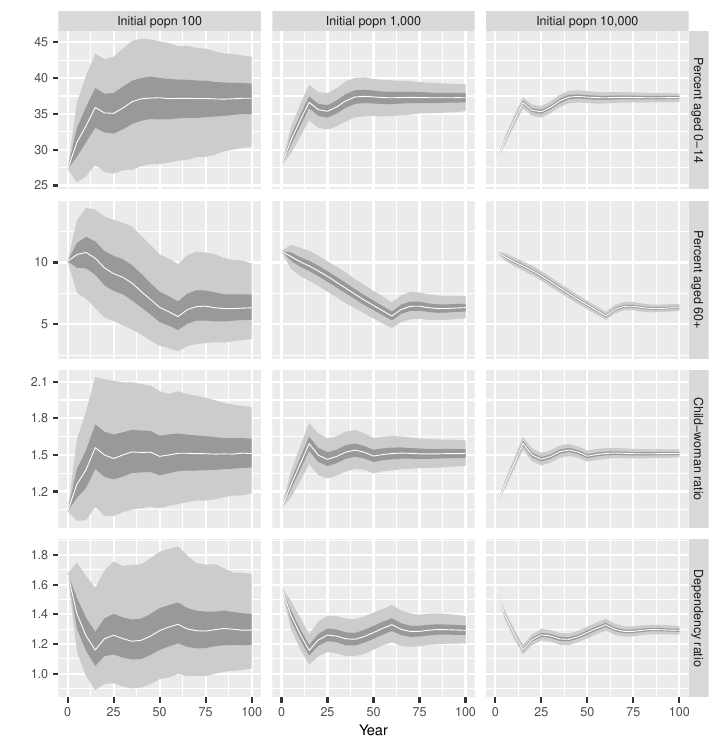}
  \caption{Distributions of summary indicators, over time, for populations with TFR of 5 and West life table level 9.}
      \label{fig:time_9_5}
\end{figure}

\end{appendices}

\end{document}